\documentclass[10pt,journal,onecolumn,compsoc]{IEEEtran}
\def\IEEEcompsocdiamondline{}

\usepackage[nointegrals]{ wasysym }
\usepackage{ amssymb }
\usepackage{ mathrsfs }
\usepackage{amsmath}
\usepackage{ dsfont }
\usepackage{kantlipsum}
\usepackage{lipsum}
\usepackage{blindtext, graphicx}
\usepackage[linesnumbered,algoruled,boxed,lined]{algorithm2e}
\usepackage{array}
\usepackage{color}
\usepackage{xcolor}
\usepackage{tikz} 

\newcolumntype{L}[1]{>{\raggedright\let\newline\\\arraybackslash\hspace{0pt}}m{#1}}
\newcolumntype{C}[1]{>{\centering\let\newline\\\arraybackslash\hspace{0pt}}m{#1}}
\newcolumntype{R}[1]{>{\raggedleft\let\newline\\\arraybackslash\hspace{0pt}}m{#1}}

\usepackage{makeidx}
\usepackage{soul}

\newcommand\copyrighttext{%
  \footnotesize This paper is a preprint (IEEE Accepted status). \textcopyright 2019 IEEE. Personal use of this material is permitted.
  Permission from IEEE must be obtained for all other uses, in any current or future 
  media, including reprinting/republishing this material for advertising or promotional 
  purposes, creating new collective works, for resale or redistribution to servers or 
  lists, or reuse of any copyrighted component of this work in other works.
  }
\newcommand\copyrightnotice{%
\begin{tikzpicture}[remember picture,overlay]
\node[anchor=south,yshift=10pt] at (current page.south) {\fbox{\parbox{\dimexpr\textwidth-\fboxsep-\fboxrule\relax}{\copyrighttext}}};
\end{tikzpicture}%
}

\begin{document}

\bstctlcite{IEEEexample:BSTcontrol} 

\title{Event Detection in Twitter Stream using Weighted Dynamic Heartbeat Graph Approach}

\author{Zafar~Saeed\\\textit{Department of Computer Science, Quaid-i-Azam University, Islamabad, Pakistan}\\\vspace{.2cm}
Rabeeh~Ayaz~Abbasi\thanks{Corresponding Author: Rabeeh~Ayaz~Abbasi (Email: rabbasi@qau.edu.pk)}\\\textit{Department of Computer Science, Quaid-i-Azam University, Islamabad, Pakistan}\\\vspace{.2cm} Muhammad~Imran~Razzak\\
\textit{Advanced Analytics Institute, University of Technology, Sydney, Australia}\\\vspace{.2cm}
Guandong~Xu\\
\textit{Advanced Analytics Institute, University of Technology, Sydney, Australia}
}

\maketitle

\copyrightnotice

\begin{abstract}

Tweets about everyday events are published on Twitter. Detecting such events is a challenging task due to the diverse and noisy contents of Twitter. In this paper, we propose a novel approach named Weighted Dynamic Heartbeat Graph (WDHG) to detect events from the Twitter stream. Once an event is detected in a Twitter stream, WDHG suppresses it in later stages, in order to detect new emerging events. This unique characteristic makes the proposed approach sensitive to capture emerging events efficiently. Experiments are performed on three real-life benchmark datasets: FA Cup Final 2012, Super Tuesday 2012, and the US Elections 2012. Results show considerable improvement over existing event detection methods in most cases.

\end{abstract}

\section{Introduction}
The proliferation of social media and blogging networks has resulted in the unprecedented growth of their users across the world. The social media stream such as Twitter provides real-time information daily about worldwide events and generates a huge amount of data, in terms of content and diversity. In recent years, there has been growing interest in detecting, analyzing and exploiting data generated by Twitter; however the analysis of the large datasets generated by Twitter is challenging due to its diverse and noisy content, and challenges associated analysis such as scalability, accuracy, and efficiency \cite{earle2012twitter,ibrahim2017tools,jarwar2017communiments}.

\begin{figure}[htbp]
 \centering
 \includegraphics [width=.7\linewidth]  {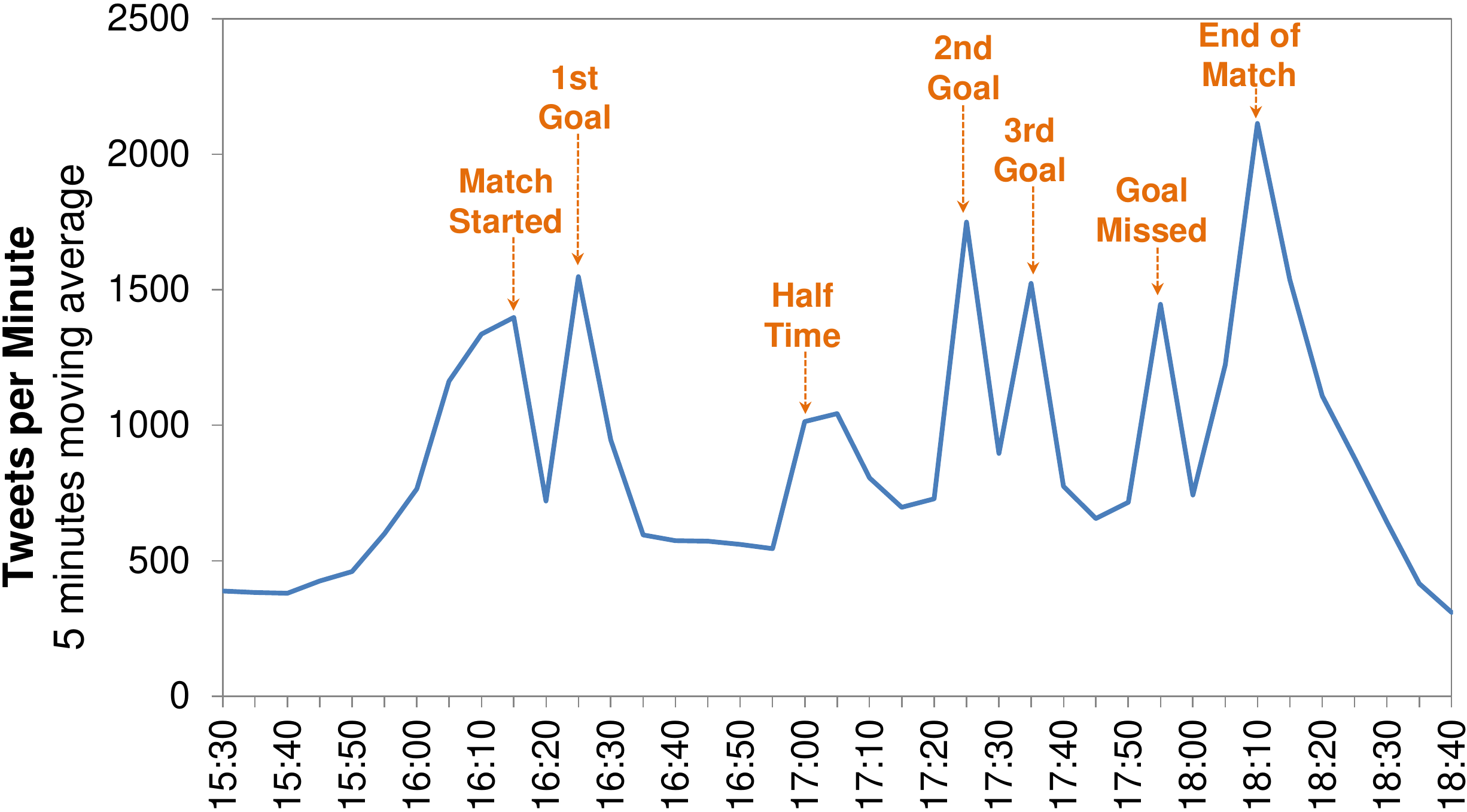} \caption{Tweets traffic at the time of Final Match of Football Association Challenge Cup, 2012 (FA Cup) between 15:30 and 18:40.}
\label{fig:TweetDistribution}
\end{figure}

Searching for event information in Twitter using keywords is a naive way, where the keyword-based search returns relevant documents matched to queried keywords. However, this approach does not guarantee that retrieved documents will represent the event accurately. Moreover, keywords selected in searches may vary from user to user, and keywords that appear in a data stream can change as the data stream evolves e.g., ``earthquake'' to ``aftershock'' or ``casualties/death''. Currently, when an event occurs, there are a few keywords that show bursty behavior (i.e., a sudden increase in usage) and the frequency of publishing micro-documents increases rapidly (Figure \ref{fig:TweetDistribution}). One simple way to find event-related information is to use trending keywords suggested by Twitter. However, trending keywords may not necessarily identify all the topics required to describe an event. Thus, sophisticated techniques are required to detect events on Twitter.

Recent work on event detection heavily relies on bursty features \cite{johansson2014global,johansson2015learning}. However, bursty features may not be able to detect events that are less popular. In this study, we develop a robust and sensitive graph-based approach. The proposed approach identifies significant changes in a text stream to detect events. This approach uses displaced temporal frequency, which is a change in frequency with respect to time, to characterize the event-related keywords and their dynamic relations.

\subsection{Problem} 

The analysis of Twitter stream, which contains diverse and noisy content, requires that a number of challenges are addressed such as accuracy, efficiency, and scalability. Micro-documents that are published to report an event have a set of related keywords that can be used to detect an event and to identify a relevant description. To identify significant information patterns, many graph-based methods have been proposed to deal with real-life data \cite{shabunina2018graph,yanardag2015deep,yanardag2015structural,yaodetecting}; however many are inadequate for the analysis of complex, dynamic and non-stationary data. They focus on bursty features that are frequent, co-occurring, and biased toward highly weighted patterns. These methods are well-placed to detect event-related topics with high frequency but ignore the dominant nature of burstiness due to which smaller events in the data are overlooked.

\subsection{Contribution}
In this study, we propose an approach Weighted Dynamic Heartbeat Graph (WDHG) to detect events. 
Once an event is detected, WDHG approach suppresses the bursty keywords at subsequent time intervals. This characteristic enables other related information to be more visible and helps in capturing new and emerging events. The proposed approach is change-sensitive and detects event-related topics efficiently. It repeatedly captures the change-patterns in the time series data of Twitter streams and highlights key occurrences.

The key contributions of our study are, therefore:
 
\begin{itemize}
\item The proposed approach (WDHG) is change-sensitive, as it subsequently suppresses high-frequency events after their first detection.

\item WDHG is computationally efficient. It detects events in polynomial time.

\item Experiments on three real-life benchmark datasets: FA Cup, Super Tuesday and the US Election demonstrate that WDHG outperforms state-of-the-art methods.

\end{itemize}

\section{Definitions} \label{sec:Preliminaries}
We start by establishing the definitions and notations used throughout the paper. We derive the following definitions from our existing work \cite{saeed2018pakdd}. \\
\noindent\textbf{Micro-document:} A micro-document (tweet) $d$ consists of a set of words $W$ posted by user $u$ at time $t$ \eqref{eq:MicroDocument}. 

\begin{equation} \label{eq:MicroDocument}
    d = (t, u, W).
\end{equation}

\noindent\textbf{Text Stream:} A text stream constitutes micro-documents \eqref{eq:TextStream}, such that $d_i$ and $d_{i-1}$ are the \textit{i-th} \textit{(i-1)-th} micro-documents published at time $\pi_1(d_i)$ and $\pi_1(d_{i-1})$,  respectively\footnote{$\pi_1(d_i)$, represents the first element ($t$) of the 3-tuple micro-document $d_i$.}, and $\pi_1(d_{i-1}) \leq \pi_1(d_i)$. 

\begin{equation} \label{eq:TextStream}
    \mathcal{D} = \{ d_1, d_2, d_3, \dots , d_n \}.
\end{equation}

Due to the short length of micro-documents, it is difficult to measure and extract meaningful information. However, systematic accumulation of micro-documents from the text stream into a super-document provides more information for detecting events. \\
\noindent\textbf{Super-document:} Let a set $\mathcal{D}$ represent a text stream consisting of all the micro-documents. A super-document $d^\rho$ is a continuous temporal accumulation of micro-documents separated at $t_a$ and $t_{a+b}$ time intervals \eqref{eq:SuperDocument} (we refer to as $t_i$ later in this paper).

\begin{equation} \label{eq:SuperDocument}
    \mathcal{D}^\rho = \{ \{d_1, d_2, \dots , d_p\} , \{d_{p+1}, \dots , d_{p+q} \} , \dots , \{ \dots , d_n\} \}.
 \end{equation}

The identity of each micro-document $d_i$ is retained while creating $k$ partitions in the text stream. A partition in $\mathcal{D^\rho}$ represents a super-document $d^\rho_i$. The aggregation strategy increases the cohesiveness among the topics when the super-document stream $\mathcal{D}^\rho$ is transformed into a graph series (see Subsection \ref{subsec:DynamicHeartbeatGraph}). Thus, a stream of super-documents consists of $k$ mutually exclusive partitions and the condition $\bigcap_{i=1}^{|\mathcal{D}^\rho|} d^\rho_i = \varnothing$ must be satisfied.
\\
\noindent\textbf{Sliding Window:} A sliding window is a moving time interval over a set of super-documents stream with a temporal coverage $\Delta t$. The data stream covered within the sliding window is processed independently to detect events and related information.

\noindent\textbf{Graph}: A graph $G_i$ consists of $G_i = (V,E,\mathscr{W},\mathscr{S})$, where $V$ represents nodes in a way that $v_i \in V$, where $V$ consists of unique words appearing in $d^\rho_i$, and $E \subseteq V \times V$ represents edges in a way that $e_k = (v_k,v_j) \land v_k \neq v_j$. $\mathscr{W}:V\rightarrow \mathds{R}$ and $\mathscr{S}:E\rightarrow \mathds{R}$ represent functions for assigning weights to nodes and edges in $G_i$ \eqref{eq:nodeWeight} and \eqref{eq:edgeWeight}, where $|d^\rho_i (v_k)|$ is the term-frequency of $v_k$ and $|d^\rho_i (v_k,v_j)|$ represents the number of co-occurrences of nodes $v_k$ and $v_j$ in a super-document $d^\rho_i$.

\begin{equation} \label{eq:nodeWeight}
    \mathscr{W}(v_k) = |d^\rho_i (v_k)|,
\end{equation}
\begin{equation} \label{eq:edgeWeight}
    \mathscr{S}(e_k) = |d^\rho_i (v_k,v_j)|.
\end{equation}

\section{Weighted Dynamic Heartbeat Graph (WDHG) Approach} \label{sec:WDHGFramework}

\begin{figure*}[htbp] 
 \centering
 \includegraphics [width=.65\textwidth] {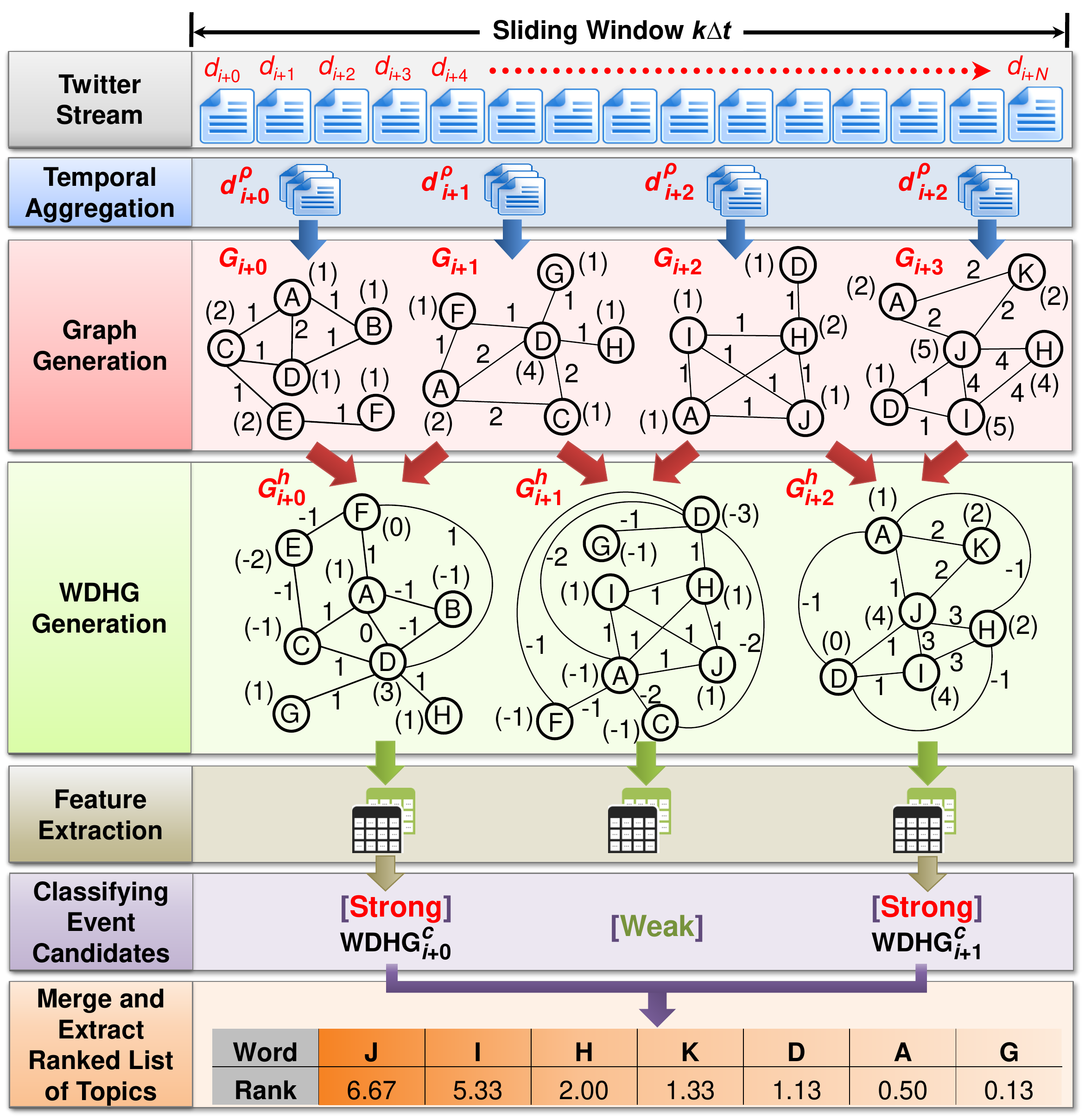}
\caption{Work-flow diagram illustrating data transformation and process starting from input as a text stream to output as a ranked list of event-related topics.}
\label{fig:Work-Flow}
\end{figure*}

Figure \ref{fig:Work-Flow} shows the workflow of the proposed approach and steps involved to transform the Twitter text stream into temporal graphs for extracting event-related topics. The data undergoes several transformations starting from the text stream as input. Micro-documents in the text stream are accumulated to generate super-documents (see Section \ref{sec:Preliminaries}). A series of graph inheriting word co-occurrence relationships from the micro-documents are generated using a set of super-documents. Furthermore, each adjacent pair of graphs is mapped onto a WDHG (see Subsection \ref{subsec:DynamicHeartbeatGraph}). Afterward, features are extracted (see Subsection \ref{subsec:DetectionModel}). A rule-based classifier labels the candidate WDHGs for event representation. To classify WDHGs as event candidates, we use aggregated centrality as a key feature. Finally, all the candidate WDHGs are merged to extract event-related topics (see Subsection \ref{subsec:TrendingTopics}). 

This study is an extension of our previous work \cite{saeed2018pakdd}. Unlike our previous work, we construct a weighted graph structure and use three different approaches for the event detection model (see Subsection \ref{subsec:DetectionModel}). We evaluate the performance of each detection model (see Subsection \ref{subsec:Results}). We extend experiments with a larger benchmark dataset (i.e., the US Election) for the evaluation. We also perform a detailed execution time analysis to determine the efficiency of the proposed approach (see Subsection \ref{subsec:ExecutionTimeAnalysis}).

\subsection{Assumptions}\label{assumption}
For detecting events, the proposed approach works with the following assumptions:
\begin{itemize}
    \item There is a significant change in the temporal frequency of words, \textit{or} new words appear between $t_{i-1}$ and $t_i$ time intervals.
    \item Significant words remain connected and result in a cohesive structure in the graph.
\end{itemize}

\subsection{WDHG Series} \label{subsec:DynamicHeartbeatGraph}
Instead of trying to compute the features directly from the text stream, we developed a flexible approach by capturing the co-occurrence relationship among the words in the form of a graph series. We create a graph series $G_i$ using the sliding window paradigm. Each node in $G_i$ is a ``word''. An edge represents the co-occurrence relationship between nodes. A graph series is a set of graphs $\mathcal{G} = \{G_1, G_2, G_3, \dots , G_{|\mathcal{D}^\rho|}\}$, where each $G_i \in \mathcal{G}$ is generated for $d^\rho_i \in \mathcal{D}^\rho$ such that $G_i$ is a labeled graph. Each graph $G_i$ in the graph series is temporally aligned with the Twitter stream and possesses the coherence relationship among the words of each micro-document $d_i$. Clique among words of each micro-document is created to increase the central tendency of topics within the graph structure.

After transforming the text stream into a graph series, we create a weighted dynamic heartbeat graph series $\mathcal{G}^h$ by linearly combining and mapping every pair of adjacent graphs ($G_{i-1}$  and $G_i$) from the graph series $\mathcal{G}$. A WDHG $G^h_i$ is a difference graph formed by combining two subsequent graphs from $\mathcal{G}$. It discriminates new and existing topics. 

\begin{figure}[htbp]
\begin{minipage}[t]{\linewidth}
    \centering
    \includegraphics [width=.55\textwidth] {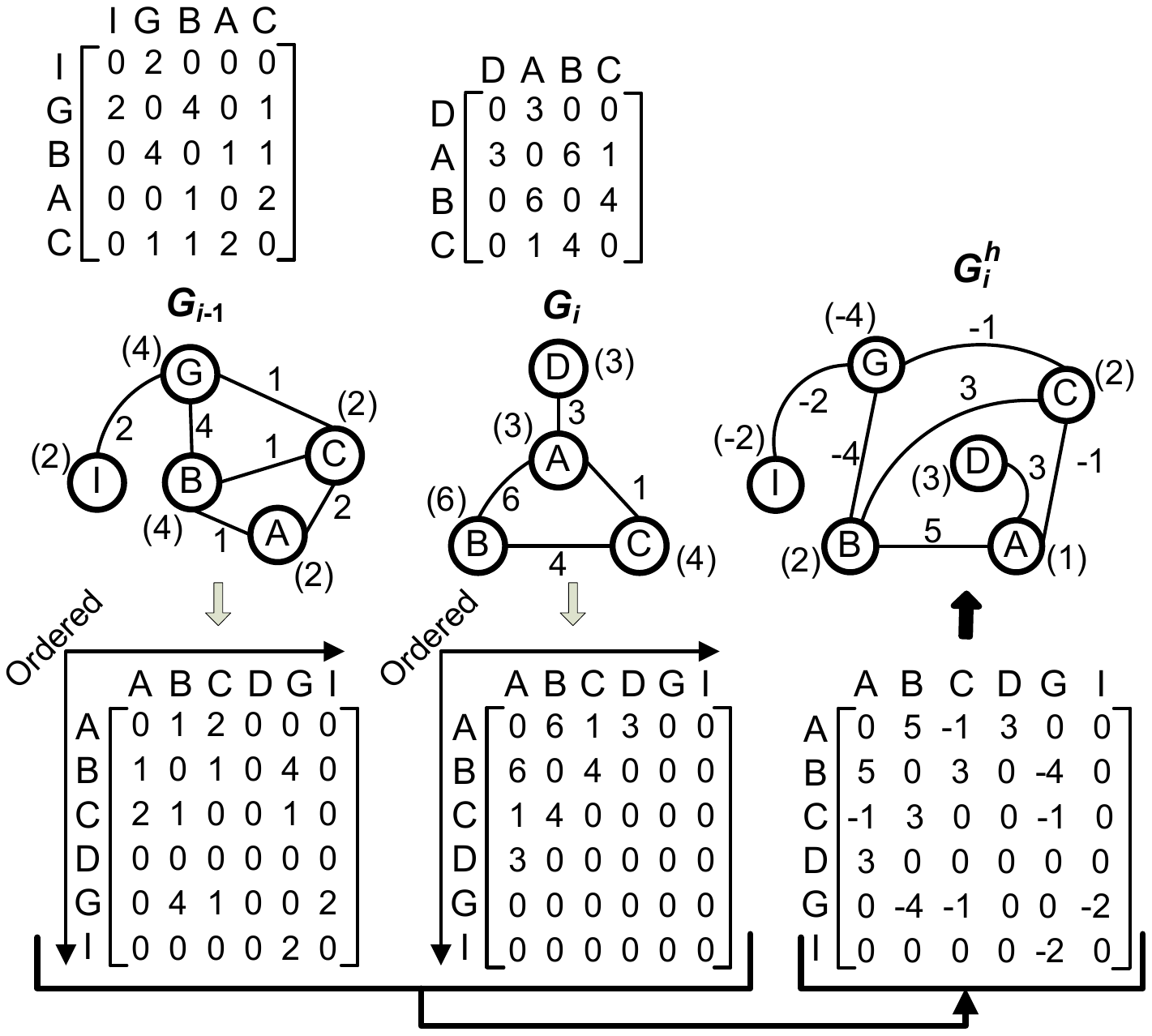}
\caption{ The example illustrates the process to generate a WDGH.}
\label{fig:WDHGexample}
\end{minipage}
\end{figure}

Transformation to WDHG is not straightforward. Adjacent graphs differ in their structures. Canonical order of the nodes in the graphs is not identical. Graphs also differ in the number of nodes. To overcome this issue, we align the dimensions of adjacent graphs ($G_{i-1}$ and $G_i$) by regenerating their adjacency matrices in equal dimensions and canonical order. The procedure of a WDHG generation is briefly described in Algorithm \ref{algo:WDHG}. Figure \ref{fig:WDHGexample} illustrates an example to generate a WDHG from two adjacent graphs $G_{i-1}$ and $G_i$. Node weights are shown in ``()''. Edge weights can be seen beside the edges.

Computational complexity improves significantly from $O(K|V|^4)$ to $O(K|V|^2)$ using above-mentioned transformation method, where $K=|\mathcal{G}|$ is the number of graphs in the graph series $\mathcal{G}$.

\begin{algorithm}
\SetKwData{VertexA}{$V^{G_i}$}
\SetKwData{VertexB}{$V^{G_{i+1}}$}
\SetKwInOut{Input}{input}
\SetKwInOut{Output}{output}
\BlankLine

\Input{$ \mathcal{G} = \{ G_1, G_2, G_3, \dots , G_{|P|}\}$ a set of temporal graphs generated against a set of super-documents $D^\rho$}
\BlankLine
\Output{$\mathcal{G}^h = \{ G^h_1, G^h_2, \dots , G^h_{|\mathcal{G}|-1}\}$}
\BlankLine
\BlankLine
\ForEach{$G_i \in \mathcal{G}$}
{
    \BlankLine
    merge set of nodes $\VertexA$ and $\VertexB$ by taking union $U$
    \BlankLine
    regenerate adjacency matrix $A$ for $G_i$ using $U$ 
    \BlankLine
    regenerate adjacency matrix $B$ for $G_{i+1}$ using $U$ 
    \BlankLine    
    calculate difference graph $G^h_i$ using $A,B,\VertexA,$ and $\VertexB$
    \label{step:heartbeatFunctionCall}
    \BlankLine    
    \BlankLine
} 

\caption{Generate Set of WDHG}\label{algo:WDHG}
\end{algorithm}

One of the inherent characteristics of WDHG is to suppress and handle bias in the data caused by the burstiness of dominating topics. Each WDHG is formulated by estimating the displaced temporal frequency of nodes and edges between each pair of graphs $G_{i-1}$ and $G_i$ to highlight other event-related information, which might also be important but less frequent at the same time. Algorithm \ref{algo:WDHG} calculates the change in the corresponding pair of graphs $G_{i-1}$ and $G_i$. It generates a WDHG $G^h_i$ by assigning new weights to all the nodes and edges. The graph series $\mathcal{G}$ is transformed into WDHG series $\mathcal{G}^h$ in a streaming fashion without affecting the temporal alignment of the data. Furthermore, these WDHGs are used to detect emerging events at precise time descriptions. The methodology of the event detection process is described in the next section.

\subsection{Event Detection} \label{subsec:DetectionModel}

The text stream has diverse contents and the heartbeat graph inherits the structural and co-occurrence relationship of a text stream. To understand our event detection model, let $\psi = G^h_i$ where $G^h_i$ is \textit{i-th} heartbeat graph for simplification. To detect the strong event candidates, we extract \textit{growth factor} and \textit{aggregated centrality} as key features from each WDHG and perform three experiments to compute corresponding heartbeat scores.

\subsubsection{Growth Factor Based:}

In the first experiment, we use growth factor $GF(\psi)$ to compute the heartbeat. $GF(\psi)$ is an accumulated score of node weights in a WDHG $\psi$ which shows the intensity of drift in topics and their popularity in the text stream. It also shows how previously observed topics are trending, in terms of popularity and if new topics are emerging at the time interval $t_i$ compared to $t_{i-1}$ \eqref{eq:growthRate}.
\begin{equation} \label{eq:growthRate}
    GF(\psi) = \sum_{k=1}^{|V^{\psi}|} \vartheta(v^\psi_k),
\end{equation}
where $\vartheta(v^\psi_k)$ is the $k^{th}$ node weight which represents the displaced temporal frequency of a word between $G_i$ and $G_{i-1}$. Heartbeat score based on growth factor is calculated as shown in \eqref{eq:HeartbeatScoreGrowthFactor}.

\begin{equation} \label{eq:HeartbeatScoreGrowthFactor}
   \mathscr{H}(\psi) =  GF(\psi).
\end{equation}

\subsubsection{Aggregated Centrality Based:}

Aggregated centrality represents the central tendency of different topics and their coherence in the WDHG $\psi$. We use aggregated topic centrality to compute the heartbeat score. Topic centrality $TC(v^\psi_k)$ expresses the central tendency of words in each WDHG $\psi$ which signifies the theme of discussion in the text stream at a certain time interval $t_i$. It is calculated using \eqref{eq:ConditionalTopicCentrality}.

\begin{equation} \label{eq:ConditionalTopicCentrality}
    TC(v^\psi_k) = \frac{\sum\limits_{i=1}^{|\varepsilon^\psi|} \pi_3(\varepsilon_i^\psi)[\pi_1(\varepsilon_i^\psi) = k \lor \pi_2(\varepsilon_i^\psi) = k)]}{|V^\psi|},
\end{equation}
where $v^\psi_k$, $\varepsilon^\psi$, and $|V^\psi|$ represent a node, indexed edge vector, and the total number of nodes in the WDHG $\psi$, respectively. $\pi_1(\varepsilon_i^\psi)$ and $\pi_2(\varepsilon_i^\psi)$ are the indexes of the nodes connected to the edge $\varepsilon_i^\psi$, and $\pi_3(\varepsilon_i^\psi)$ is the weight of the edge. The centrality scores of all nodes that contain at least one positive edge in the WDHG $\psi$ are accumulated to calculate the aggregated centrality score $AC(T^\psi)$ using \eqref{eq:IndexVectorConversion} and \eqref{eq:AggregatedCentrality}, where $T^\psi$ is a set of indexes of those nodes that are connected to at least one positive edge.

\begin{equation} \label{eq:IndexVectorConversion}
    T^\psi = \bigcup\limits_{i=1}^{|\varepsilon^\psi|}\left(\pi_1(\varepsilon_i^\psi) \cup \pi_2(\varepsilon_i^\psi)\right),
\end{equation}

\begin{equation} \label{eq:AggregatedCentrality}
    AC(T^\psi) = \sum\limits_{k=1}^{|T^\psi|}TC(v^\psi_{T^\psi_k}).
\end{equation}

The indexed edge vector $\varepsilon^\psi$ is used to calculate aggregated centrality. It contains only those edges that have positive weights. Due to the initial assumption (see Section \ref{sec:WDHGFramework}) in the proposed approach, all the negative edges are dropped. It improves the centrality of newly emerging topics in the graph structure with respect to existing ones. It also reduces the number of passes significantly, thereby improving execution time. A high aggregated centrality score of a WDHG $\psi$ shows that the keywords are coherent and emerging topics are concurrently reported in the Twitter stream at a certain time interval $t_i$. Heartbeat score based on aggregated centrality is calculated using \eqref{eq:HeartbeatScoreAggregatedCentrality}.

\begin{equation} \label{eq:HeartbeatScoreAggregatedCentrality}
      \mathscr{H}(\psi) =  AC(T^\psi).
\end{equation}

\subsubsection{Aggregated Centrality and Growth Factor Based:}
We also combine aggregated centrality and growth factor by multiplying both features to calculate heartbeat score \eqref{eq:HeartbeatScorecombine}.

\begin{equation} \label{eq:HeartbeatScorecombine}
      \mathscr{H}(\psi) =  GF(\psi) \times AC(T^\psi).
\end{equation}

\subsubsection{Emerging Event Identification:} 

To detect event candidates, two labels \textit{Strong} or \textit{Weak} are assigned to each WDHG $\psi$. \textit{``Strong''} means WDHG $\psi$ contains emerging event descriptions and \textit{``Weak''} means WDHG $\psi$ is insignificant. A rule-based classification function $Est(\psi)$ \eqref{eq:ClassificationFunction} estimates and assigns class labels to each WDHG $\psi \in \mathcal{G}^h$.

\begin{equation} \label{eq:ClassificationFunction}
    Est(\psi)= 
        \begin{cases}
            \text{``Strong''},& \text{if } \mathscr{H}(\psi) \geq \theta_{(k\Delta t).}\\
            \text{``Weak''},              & \text{otherwise.}
        \end{cases}
\end{equation}
Here, $\theta$ is a dynamic threshold computed based on heartbeat scores of WDHGs. It sets an optimum value for classification function $Est(\psi)$ \eqref{eq:ThetaCalculation} in the sliding window $k\Delta t$.

\begin{equation} \label{eq:DHGCountInTimeSlotForTheta}
    \mathcal{N} = \frac{\Delta t}{t_i},
\end{equation}

\begin{equation} \label{eq:MeanForTheta}
    \varpi = \frac{\sum\limits_{i=k}^{\mathcal{N}+k} (HB(\psi))}{\mathcal{N}},
\end{equation}

\begin{equation} \label{eq:ThetaCalculation}
    \theta_{(k\Delta t)} = \varpi + \omega \sqrt{\frac{\sum\limits_{i=k}^{\mathcal{N}+k}(HB(\psi) - \varpi)^2}{\mathcal{N}}},
\end{equation}
where $\Delta t$ and $t_i$ are the temporal coverages of the sliding window and super-document $d^\rho_i$, respectively. $\mathcal{N}$ is the number of WDHGs in the sliding window \eqref{eq:DHGCountInTimeSlotForTheta}. $\varpi$ is the average heartbeat score within a sliding window \eqref{eq:MeanForTheta}. $\omega$ is the adjustment parameter that deals with the data dispersion. $k$ is the index of the first WDHG in the sliding window under consideration, and $HB(\psi)$ is the heartbeat score of WDHG $\psi$.

\subsection{Trending Topics} \label{subsec:TrendingTopics}

\begin{figure}[htbp]
\begin{minipage}[t]{\linewidth} 
\centering
\includegraphics [width=.55\textwidth] {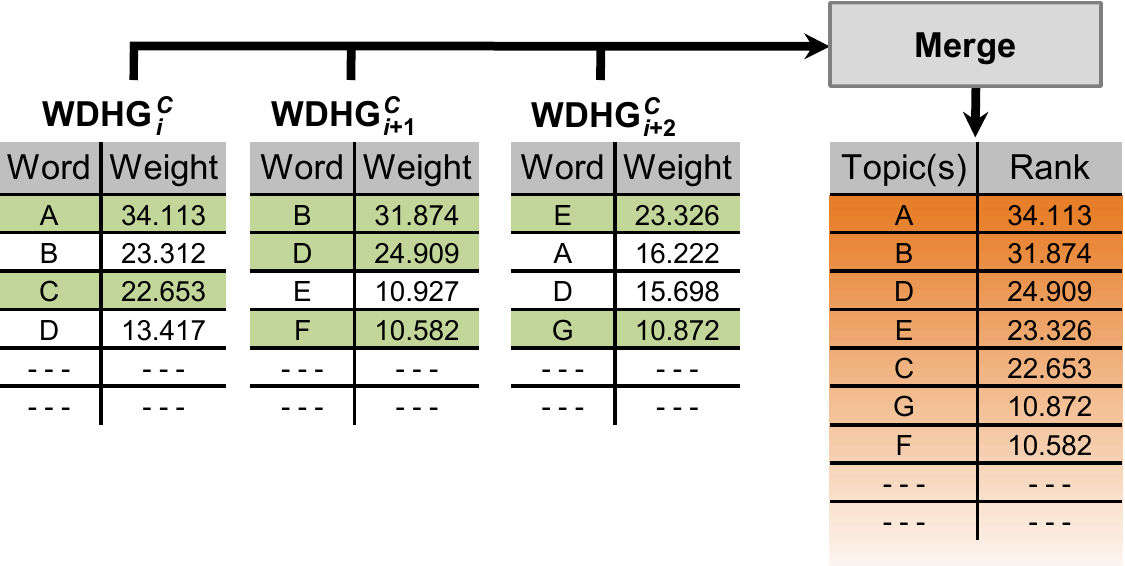}
\caption{The example illustrates the unification of multiple candidate WDHGs in a sliding window to produce a final list of topics. Each copy that has the highest weight among its duplicates is selected for the final ranked list (as highlighted in the figure).}
\label{fig:MergeCandidateWDHGs}
\end{minipage}
\end{figure}

Multiple candidate WDHGs within a sliding window can have duplicate words. We select unique words that have the highest weights among candidate graphs to extract the event-related topics without compromising performance. The weight for each word is calculated by fusing degree centrality and displaced temporal frequency scores. The example shown in Figure \ref{fig:MergeCandidateWDHGs} further elaborates the process of merging multiple candidate WDHGs, where $WDHG^c_i$ is the \textit{i-th} candidate graph in the sliding window under process. In the sliding window $k\Delta t$, a ranked list of topics is extracted from all candidate WDHGs and classified as \textit{``Strong"}. Topic ranks are calculated using \eqref{eq:TopicRanking}.

\begin{equation} \label{eq:TopicRanking}
    Rank(v^{\psi}_k) = \mathscr{C}(v^\psi_k) \times \mathscr{W}(v^\psi_k).
\end{equation}

\begin{figure}[htbp]
\begin{minipage}[t]{\linewidth}
    \centering
    \includegraphics [width=.5\textwidth] {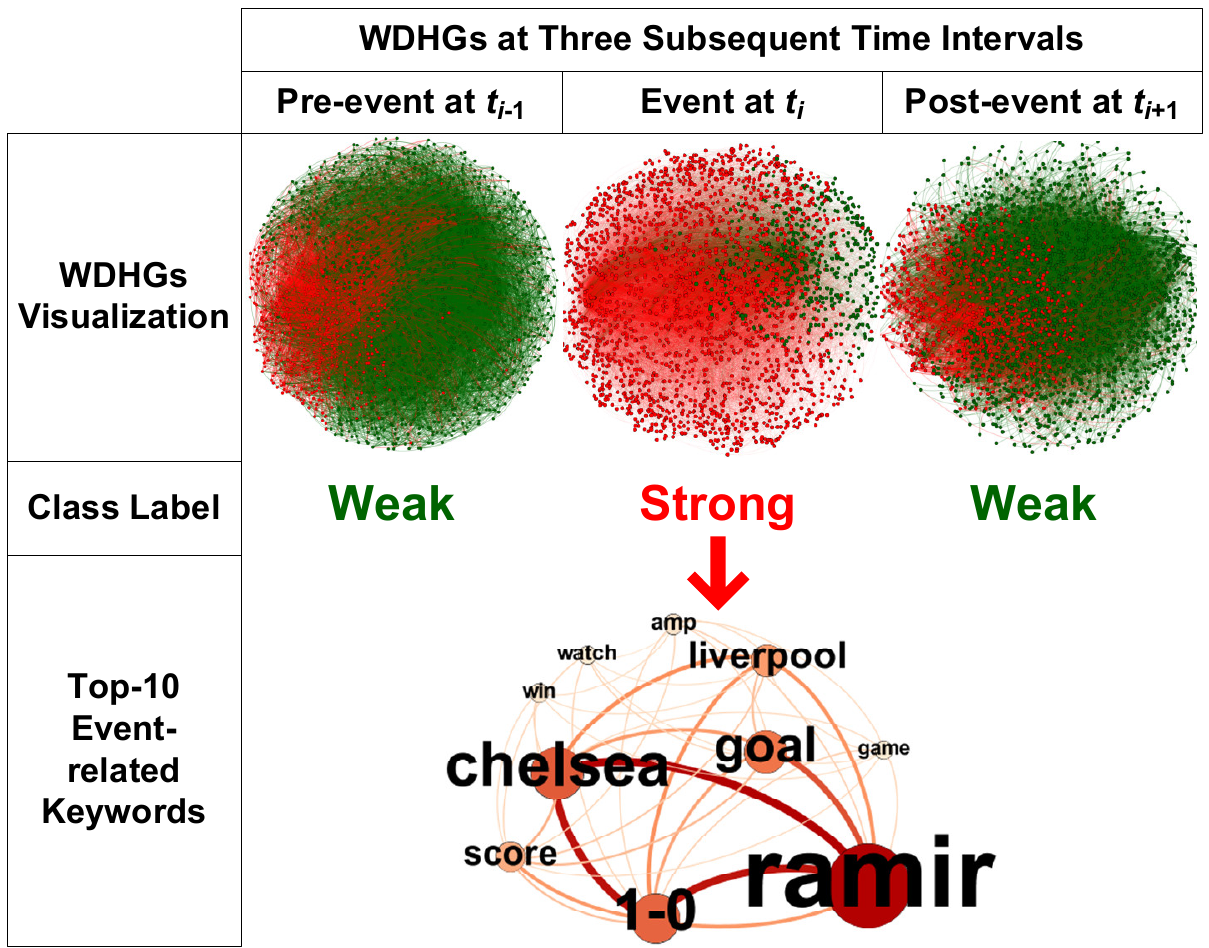} 
\caption{ Visualization of three WDHGs $G^h_{i-1}$, $G^h_i$, and $G^h_{i+1}$ at time $t_{i-1}$, $t_i$, and $t_{i+1}$, respectively using the FA Cup dataset. A significant event ``Goal'' occurred at interval $t_i$. The nodes and edges with positive and negative weights are shown in red and green, respectively. A large portion of the graph is affected when an event occurs. It shows that WDHG is hyper-sensitive to detect newly emerging topics.}
\label{fig:visualization}
\end{minipage}
\end{figure}

Figure \ref{fig:visualization} shows the visualization of three consecutive WDHGs with their class labels and the top ten keywords from an event (i.e., Goal) in the FA Cup dataset. The event detection model is described in Algorithm \ref{algo:EventDetection}. It consists of feature extraction, event candidate detection, and finally extraction of an event-related ranked list of topics.

\begin{algorithm}
\SetKwInOut{Input}{Input}
\SetKwInOut{Output}{Output}
\BlankLine

\Input{ $ \mathcal{G}^{h(k\Delta t)} - $ Set of WDHGs within the sliding window $k\Delta t$}
\BlankLine
\Output{$\mathcal{L} - $ List of ranked topics}
\BlankLine
\BlankLine

\ForEach{$G^h_i \in \mathcal{G}^{h(k\Delta t)}$}
{
    \BlankLine
    calculate growth factor $GF_i$ using \eqref{eq:growthRate}
    \BlankLine
    calculate aggregated centrality $AC_i$ using \eqref{eq:AggregatedCentrality}
    \BlankLine
    calculate heartbeat score $HB_i$ using \eqref{eq:HeartbeatScorecombine}
}
\BlankLine
calculate $\theta_{(k\Delta t)}$ using \eqref{eq:ThetaCalculation} for $\mathcal{G}^{h(k\Delta t)}$
\BlankLine

\ForEach{WDHG $G^h_i \in \mathcal{G}^{h(k\Delta t)}$}
{
    \BlankLine
    assign binary class label $C$ to $G^h_i$ for the corresponding heartbeat score $HB_i$ using \eqref{eq:ClassificationFunction}
    \BlankLine
    \If(){$C =  $``Strong''}
    {
        \BlankLine
        assign weight to each keyword in $G^h_i$ using \eqref{eq:TopicRanking}
        \BlankLine

        merge keywords from $G^h_i$ into the topic list $\mathcal{L}$ 
        
        \BlankLine
        remove duplicates from $\mathcal{L}$ while keeping keywords that have maximum weight
        
    }
    
}

\BlankLine
sort $\mathcal{L}$ based on topic ranking
    
\caption{Event Detection Algorithm}\label{algo:EventDetection}
\end{algorithm}

\section{Experiments and Evaluation} \label{sec:ExperimentAndResults}

We propose three event detection approaches. These approaches are based on the measures growth factor and topic centrality and their combination (see Subsection \ref{subsec:DetectionModel}). To test and evaluate these three approaches, we first compare their performances with each other; then the best approach is selected and compared with nine state-of-the-art methods (see Subsection \ref{subsec:Results}). We perform experiments on three benchmarks: FA Cup, Super Tuesday and the US Election datasets. An existing evaluation framework is used for comparative analysis \cite{aiello2013sensing}.

\subsection{Parameters} \label{subsec:Parameters}

Three parameters (i.e., $\Delta t$, $t_i$, and $\omega$) are used for optimization. $\Delta t$ is the temporal coverage of the sliding window that contains a batch of data to be processed independently. As per ground-truth, the temporal coverage $\Delta t$ of sliding window is set to one minute, one hour, and ten minutes for FA Cup, Super Tuesday, and the US Election datasets, respectively. The small size of the micro-document in the Twitter stream did not yield useful information; therefore, micro-documents are accumulated together that appear within a specific period to form a super-document. The parameter $t_i$, a temporal coverage of super-documents, is set to one minute, ten minutes, and one minute for the FA Cup, Super Tuesday, and the US Election datasets, respectively. $\omega$ is the adjustment parameter that deals with the data dispersion. Its values of 1, 0.6, and 0.6 are used for FA Cup, Super Tuesday and the US election datasets, respectively.

\subsection{Evaluation Measures} \label{subsec:EvaluationMeasure}

To evaluate the performance, we compare the results with the ground truth using two metrics:
\begin{itemize}
    \item \textit{Topic-Recall@K} (T-Recall): Fraction of topics detected successfully from the ground truth
    \item \textit{Keyword-Precision@K} (K-Precision): Fraction of keywords detected successfully from the top-K retrieved keywords
\end{itemize}
As there are multiple topics in the benchmark datasets (see Table \ref{tbl:DatasetStats}), the final evaluation metrics are calculated by micro-averaging the \textit{T-Recall} and \textit{K-Precision} of individual topics. The ground truth is created based on the events reported in the mainstream media. Topic precision cannot be used for the evaluation as the text stream contains several newsworthy events which are not included in the ground truth \cite{aiello2013sensing}.

\subsection{Dataset} \label{subsec:DataCollection}

The experiments are conducted on three well-known benchmarks: FA Cup, Super Tuesday, and the US Election datasets \cite{aiello2013sensing}. Many recent studies \cite{adedoyin2016rule,choi2019emerging,elbagoury2015exemplar,ibrahim2017tools,nguyen2017real,papadopoulos2014snow,prabandari2017comparative} use these benchmarks to evaluate the performance of their approaches. 
The details of the three datasets are given in Table \ref{tbl:DatasetStats}.

\begin{table}[htbp]
\centering
\caption{Datasets detail and temporal coverage}
\label{tbl:DatasetStats}
\begin{tabular}{l|c|c|c|}
\cline{2-4}
 & Temporal Coverage & No. of Tweets & Total Topics \\ \hline
\multicolumn{1}{|l|}{FA Cup} & 6 hours & 124,524 & 13 \\ \hline
\multicolumn{1}{|l|}{Super Tuesday} & 24 hours & 540,241 & 22 \\ \hline
\multicolumn{1}{|l|}{US Election} & 36 hours & 2,335,105 & 64 \\ \hline
\end{tabular}
\end{table}

The ``FA Cup'' dataset contains tweets posted during the final match of the Football Association Challenge Cup held on May 5, 2012. FA Cup is one of the oldest football competitions with a huge fan base. The match was played between the Chelsea and Liverpool teams. Chelsea won the Cup 2-1, Ramirez and Drogba scored each of the two goals from the winning team. The only goal from Liverpool was scored by Carrol. The ground truth for the FA Cup dataset comprises 13 topics, including kick-off, goals, half-time, fouls, bookings, and the end of the match.

The ``Super Tuesday'' dataset consists of tweets posted during the US primary elections, which were held on the first Tuesday of March 2012 in ten US states. The ground truth comprises 22 topics, which represents the key moments of the elections and projections of the voting results in different states.

The US Election dataset contains tweets posted during the United States presidential election of 2012 which was held on November 6, 2012. The ground truth consists of 64 topics. The topics were related to the outcomes of the presidential election, derived from mainstream media.

\subsection{Ground Truth}

The ground-truth comprised several event-related topics as shown in Table \ref{tbl:DatasetStats}. Each topic in the ground-truth is expressed by a set of keywords which are further divided into mandatory and optional categories. To detect a ground-truth topic, it is essential to detect all mandatory keywords; however optional words are more expressive in terms of event description and are used collectively in evaluating keyword precision along with mandatory keywords.

\subsection{Data Pre-processing}

Users are free to write tweets in their way. A tweet might contain user mentions, hash-tags, or URLs along with its contents. It needs to be pre-processed to reduce noise. The datasets used in the study undergo the following pre-processing steps to improve their content quality:

\begin{itemize}
    \item Duplicate tweets are removed.
    \item Tweet characters are converted to lower case, and all the special characters are removed.
    \item Tokenization is performed to separate all words using white-spaces. Then, stop words and common words are removed.
    \item Words that consist of less than three letters are removed.
\end{itemize}

\subsection{Results} \label{subsec:Results}

A taxonomy of event detection techniques is proposed in a recent study \cite{ibrahim2017tools}. Ibrahim et al. explored and classified event detection techniques into five major categories. 1) Probabilistic Models, 2) Clustering, 3) Frequent Pattern Mining, 4) Matrix Factorization, and 5) Exemplar-based. We consider some of the recent and state-of-the-art approaches from each of the categories to compare and evaluate the performance of the proposed approach. Since the proposed approach is graph-based, a Graph-feature pivot method is also included in the baselines.

\begin{figure*}[htbp]
\centering
\includegraphics [width=.8\linewidth] {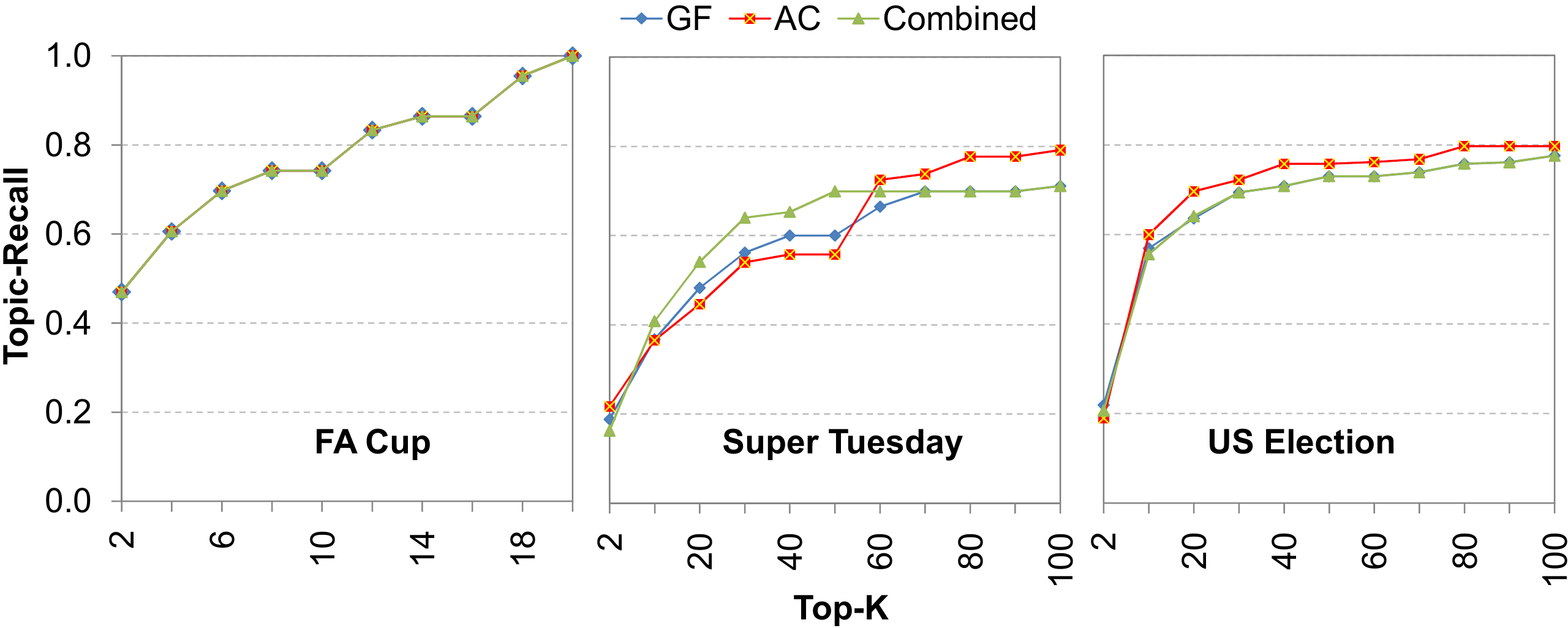} 
\caption{Comparison of event detection methods (i.e., Growth Factor (GF), Aggregated Centrality (AC), and Combined) for FA Cup, Super Tuesday, and the US Election datasets.}
\label{fig:TopicRecallall}
\end{figure*}

We perform a two-step evaluation. First, we compare the performance of our three event detection methods (growth factor, aggregated centrality and a combination of growth factor with aggregated centrality) as shown in Figure \ref{fig:TopicRecallall}. Results show that event detection based on aggregated centrality is better than growth factor or a combination of both.

Second, we compare the performance of the winner approach (i.e., aggregated centrality) as the event detection method, and perform a comparative analysis with the following state-of-the-art methods:

\begin{itemize}
    \item \textit{Probabilistic Model} - Latent Dirichlet Allocation (LDA) \cite{teh2007collapsed}
    \item \textit{Clustering} - Document-pivot (Doc-p) \cite{petrovic2010streaming}, BN-gram \cite{aiello2013sensing}
    \item \textit{Frequent Pattern Mining} - Soft Frequent Pattern Mining (SFPM) \cite{aiello2013sensing}
    \item \textit{Matrix Factorization} - SVD-KMean \cite{nur2015combination}, SNMF-Orig, SNMF-KL \cite{prabandari2017comparative}
    \item \textit{Exemplar-Based} - Exemplar \cite{elbagoury2015exemplar}
    \item \textit{Graph-based} - Graph-based Feature-pivot (GFeat-p) \cite{o2010tweetmotif}
\end{itemize}

\begin{figure}[htbp]
\centering
\includegraphics [width=.55\linewidth] {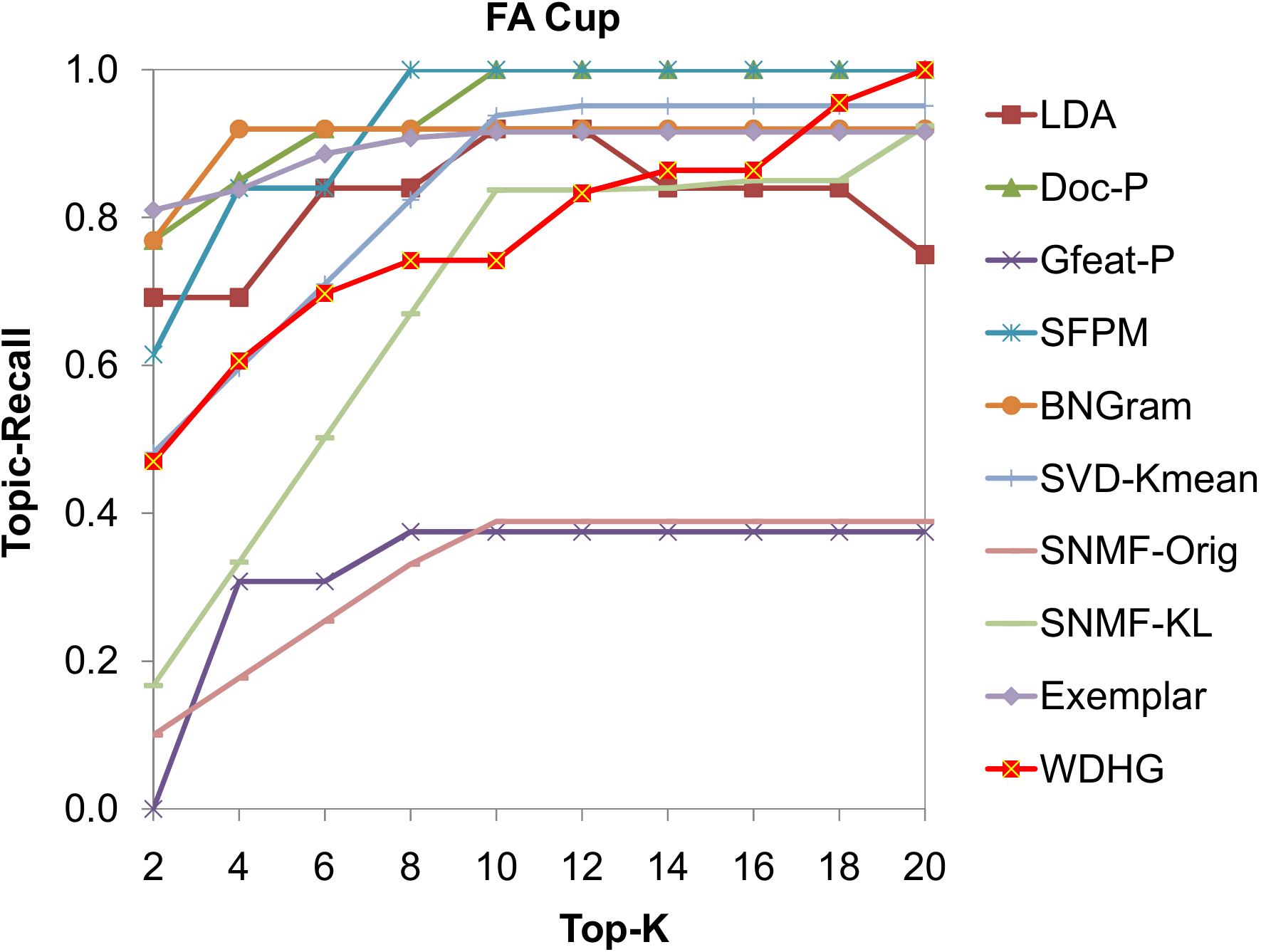} 
\caption{Comparison with baseline methods for the FA Cup dataset.}
\label{fig:TopicRecall-FA}
\end{figure}

For the FA Cup dataset, Figure \ref{fig:TopicRecall-FA} shows the results for \textit{T-Recall} at $K = \{2, 4, 6, \dots , 20$\}. For all the approaches, the results for the FA Cup are the best among all the three datasets due to short duration and high popularity of the football match. Users posting tweets for such events are focused and consistent. Therefore, topics that appeared in the data are less diverse and easier to detect than in Super Tuesday and the US Election datasets. For smaller values of $K$, WDHG does not perform well; however, it quickly gains the maximum \textit{T-Recall} and detects all the ground truth topics at $K = 20$ as shown in Figure \ref{fig:TopicRecall-FA}.

\begin{figure}[htbp]
\centering
\includegraphics [width=.55\linewidth] {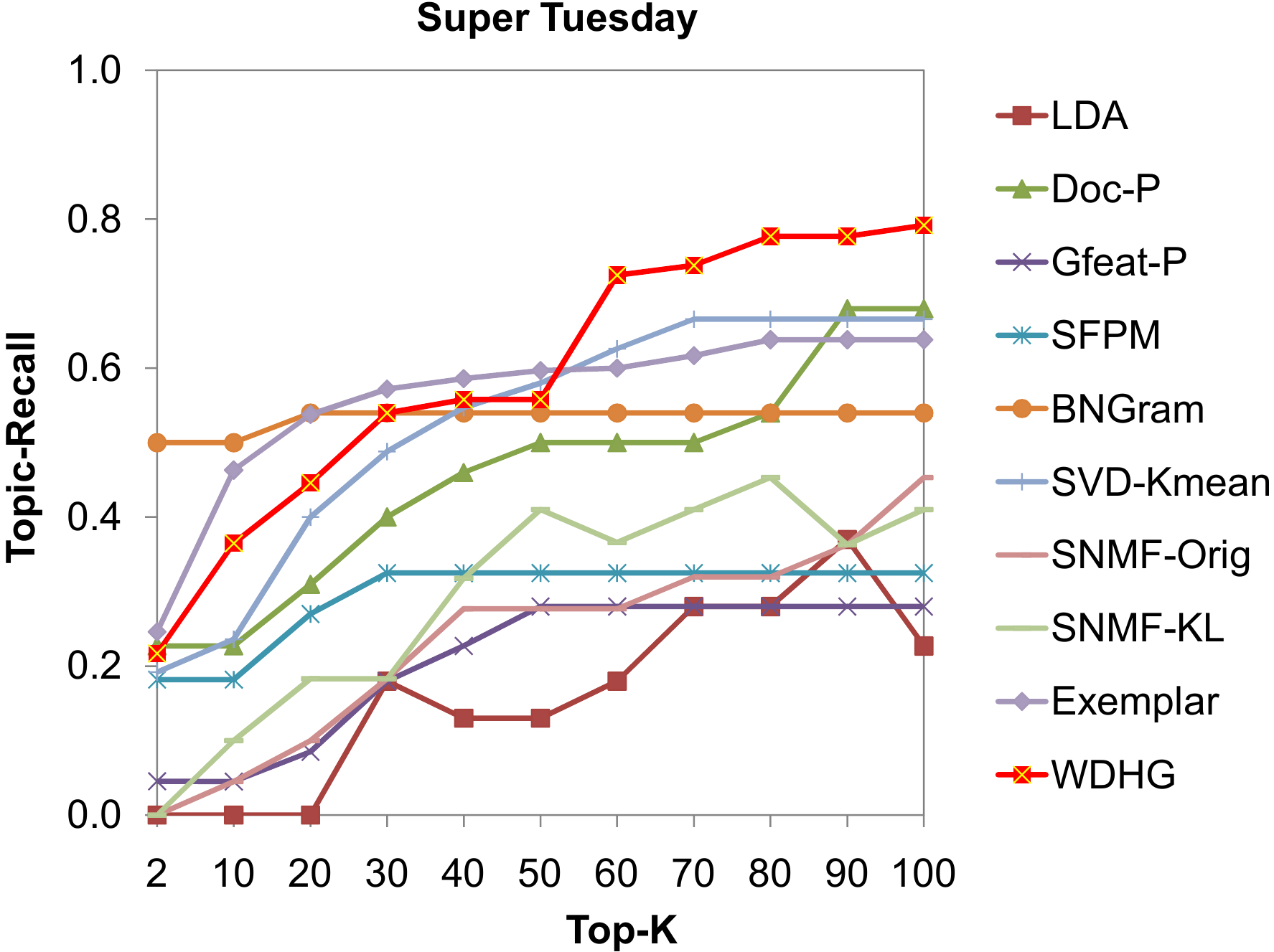} 
\caption{Comparison with baseline methods for the Super Tuesday dataset.}
\label{fig:TopicRecall-ST}
\end{figure}

Conversely, the Super Tuesday and the US Election datasets have more diversity in their text streams due to greater temporal coverage compared to the FA Cup. The fundamental characteristic of the proposed method is to sense a periodic change in the text stream and to extract coherent topics from the word-graph (see Section \ref{sec:WDHGFramework}). Therefore, the proposed method outperforms all other baselines at $K > 50$ for the Super Tuesday dataset as shown in Figure \ref{fig:TopicRecall-ST}.

\begin{figure}[htbp]
\centering
\includegraphics [width=.55\linewidth] {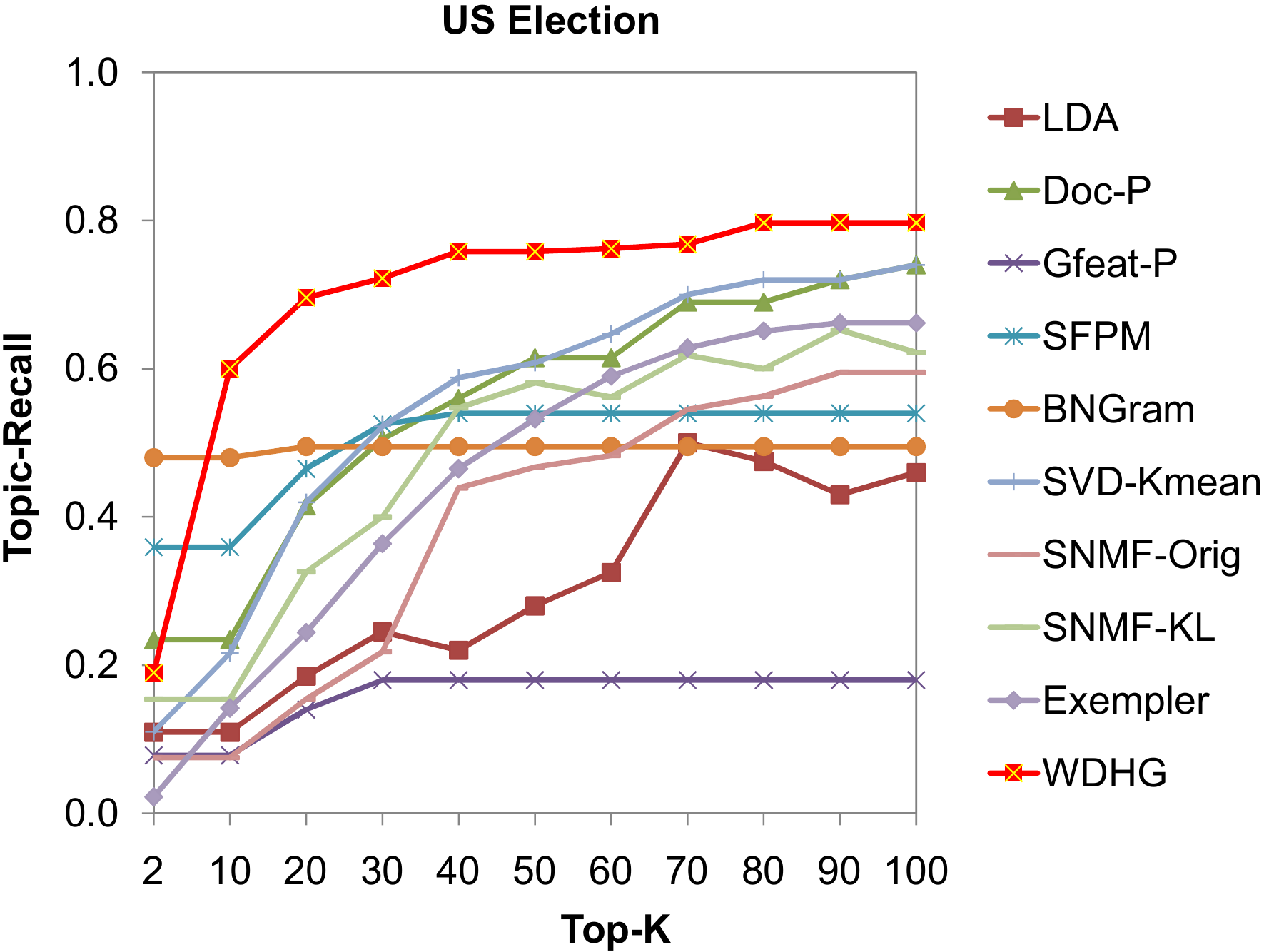} 
\caption{Comparison with baseline methods for the US Election dataset.}
\label{fig:TopicRecall-US}
\end{figure}

The US Election is one of the largest datasets in terms of temporal coverage which spans over 36 hours including sixty 64 topics. The proposed WDHG method significantly outperformed all the baseline methods at $K > 2$ as shown in Figure \ref{fig:TopicRecall-US}.

Table \ref{tab:K_Prec} shows the results of \textit{K-Precision} at $K = 2$. The proposed WDHG method combines the scores of the displaced temporal frequency and the topic centrality in the word-graph; therefore it detects even-related keywords with high precision for all three datasets. The proposed WDHG method is an effective approach and performed consistently better compared to existing methods in terms of performance and accuracy.

\begin{table}[htbp]
  \centering
  \caption{ Comparison with baseline methods for Keyword-Precision@2 for FA Cup, Super Tuesday, and the US Election datasets. Best results for each dataset are in bold face.}
  \begin{tabular}{|l|c|c|c|}
\hline
\textbf{ Method } & \textbf{ FA Cup } & \textbf{ Super Tuesday } & \textbf{ US Election } \\ \hline
    LDA   & 0.164 & 0.000  & 0.165 \\
    Doc-P & 0.337 & 0.511 & 0.401\\
    Gfeat-P & 0.000     & 0.375 &0.375 \\
    SFPM  & 0.233 & 0.471 & 0.241 \\
    BNGram & 0.299 & 0.628 & 0.405 \\
    SVD-Kmean & 0.242 & 0.367 & 0.300 \\
    SNMF-Orig & 0.330 & 0.241 & 0.241 \\
    SNMF-KL & 0.242 & 0.164 & 0.164 \\
    Exemplar & 0.300 & 0.485 & 0.391 \\
    WDHG   & \textbf{0.545} & \textbf{0.750} & \textbf{0.423} \\
    \hline
    \end{tabular}%
  \label{tab:K_Prec}%
    
\end{table}
We observe that whenever the text stream starts to change and deviate from its current trending topics, the proposed WDHG method quickly detects the emerging event due to its sensitivity towards the dynamic nature of the text stream. 

It also appears that user participation increases when an event occurs. The user participation feature is not used in the proposed WDHG method; however, it can be useful to improve the detection model further. We are planning to use it in future studies.

\subsection{Computational Complexity}

We reduce the time complexity of event candidate detection from $O(|V|^2)$ to $O(N^2)$ by transforming WDHG into vector-space $\varepsilon$ (see Subsection \ref{subsec:DynamicHeartbeatGraph}), where $V = Max(|V^{\psi}|)$, $N = Max(|\epsilon_i|)$, and $N^2 \ll |V|^2$. Considering a worst case, $O(|V|^2) \equiv O(N^2)$ if and only if, WDHG $\psi$ is a complete graph; however, the occurrence of such scenarios is quite rare because each WDHG $\psi \in \mathcal{G}^h$ is sparse.

\subsection{Execution Time Analysis} \label{subsec:ExecutionTimeAnalysis}

Twitter generates a great number of micro-documents within a short interval, and frequency of publishing micro-documents increases rapidly when a significant event occurs as shown in Figure \ref{fig:TweetDistribution}. Therefore, we evaluate the execution time of event detection methods to analyze how quickly these methods detect event-related topics with top-k keywords. We conduct experiments on a machine having a CPU Intel Core i5-3210M (2.5 GHz) processor and 16GB-DDR3 memory. Usually, the primary objective of event detection methods is to produce results with high accuracy and later achieve a better computational time. Therefore, the methods SNMF-Orig, SNMF-KL, LDA and Gfeat-P with inferior performances (see Figures \ref{fig:TopicRecall-FA}, \ref{fig:TopicRecall-ST}, and \ref{fig:TopicRecall-US}) are excluded from the execution time analysis.

\begin{figure}[htbp]
\begin{minipage}[t]{\linewidth}
\centering
\includegraphics [width=.75\textwidth] {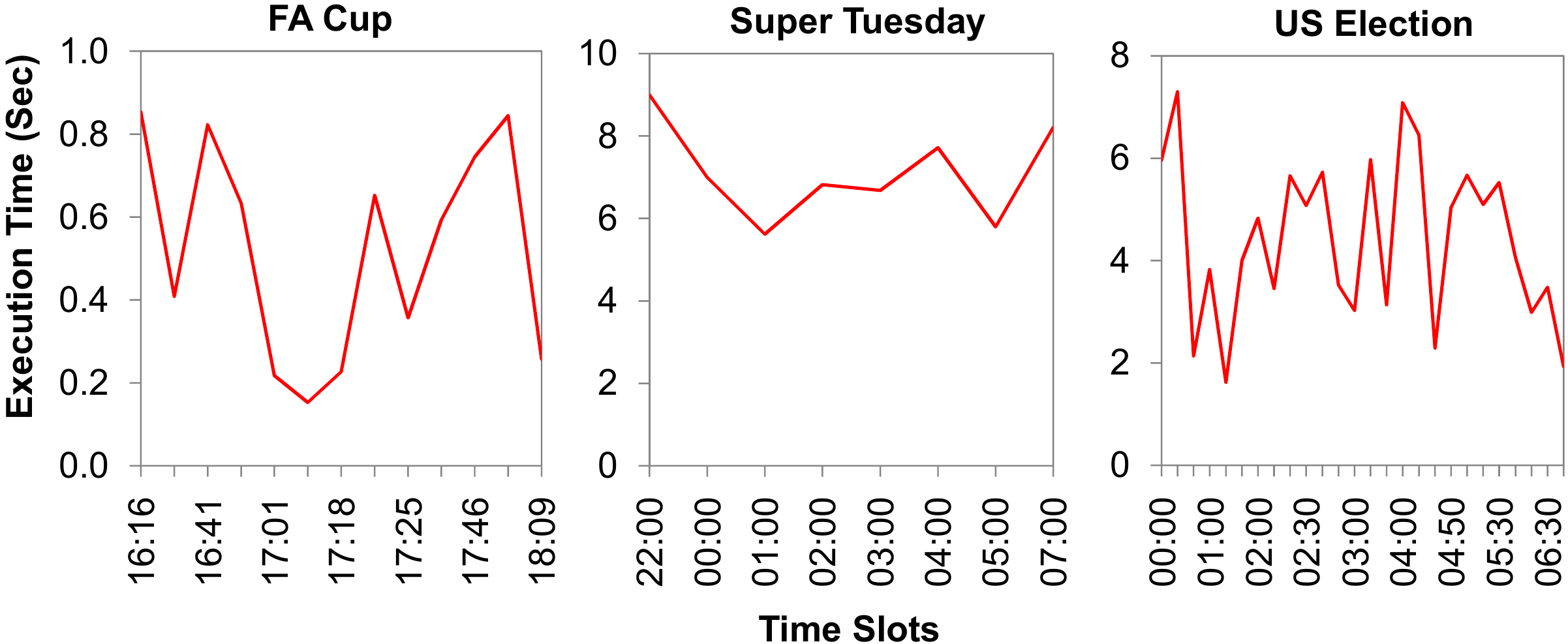}
\caption{Execution time of the proposed approach for event candidate sliding windows of FA Cup, Super Tuesday, and the US Election datasets.}
\label{fig:ExecutionTime-SlidingWindows}
\end{minipage}
\end{figure}

\begin{figure}[htbp]
\begin{minipage}[t]{\linewidth}
\centering
\includegraphics [width=.75\textwidth] {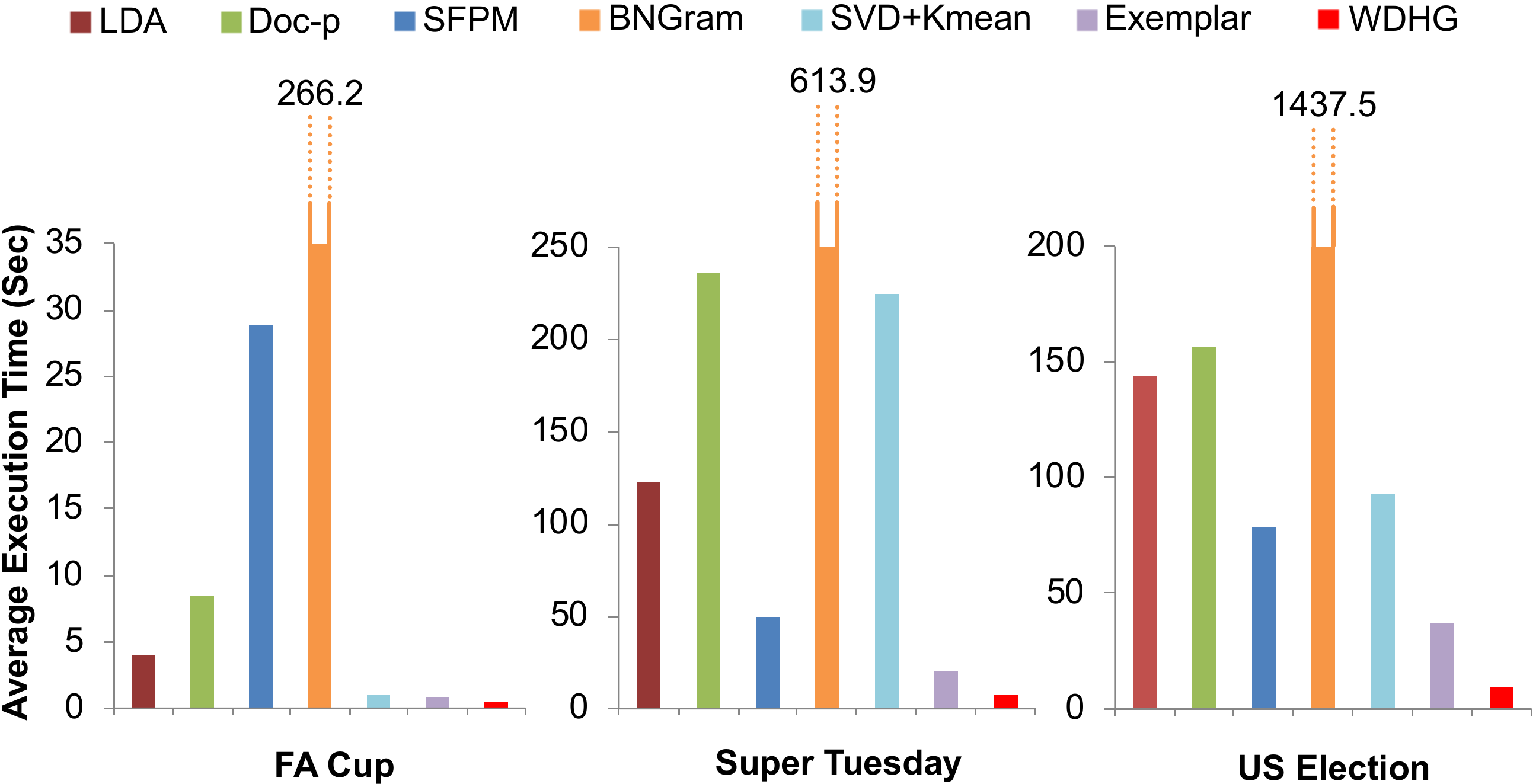}
\caption{Comparison of average execution time to produce top 20, 100, and 100 keywords for FA Cup, Super Tuesday, and the US Election datasets, respectively.}
\label{fig:ExecutionTime-Top-K-Results}
\end{minipage}
\end{figure}

Figure \ref{fig:ExecutionTime-SlidingWindows} shows the time required to process each event candidate sliding window by WDHG method. Figure \ref{fig:ExecutionTime-Top-K-Results} shows the average processing time required to produce the top 20, 100, and 100 keywords in the FA Cup, Super Tuesday, and the US Election datasets, respectively.

The average execution times (in seconds) for the proposed approach to detect event-related topics are 0.46, 7.24, and 9.52 for the FA Cup, Super Tuesday, and the US Election datasets, respectively, which is better than all of the baseline methods. BNGram has the worst execution times (266.25, 613.97, and 1437.50) for the FA Cup, Super Tuesday, and the US Election datasets, respectively. Exemplar has the second best execution times (0.88, 20.12, and 37.25) for the FA Cup, Super Tuesday, and the US Election datasets, respectively. Comparison of the execution times with Exemplar shows that the proposed approach is 47\%, 64\%, and 74\% faster for the FA Cup, Super Tuesday, and the US Election datasets, respectively.

\section{Related Work}\label{section:Relatedwork}

Impressive efforts have been made to detect events in social streams. The detection of emerging events involves the identification of trending topics related to the event by monitoring and processing the text stream. Existing event detection techniques include bursty detection, topic model and clustering algorithms. 

In order to identify significant  keywords representing events, several research studies \cite{li2012tedas,mathioudakis2010twittermonitor,nguyen2015real,shamma2011peaks,yang2011patterns} have used frequency signals while processing the text stream. Keywords that have high frequency/burst are retained and further processed to segregate the information which is later used to identify the occurrence of events. Mathioudakis et al. identified the events based on a sharp increase in keyword frequency in specified time intervals \cite{mathioudakis2010twittermonitor}. However, detection based on the intensity of frequency could not distinguish different events that shared the same keywords in the bursty flow. To find abnormal spikes in keyword-based frequency signals, He et al. used the Discrete Fourier Transformation (DFT) method to group keywords based on features extracted from periodicity and strength of the power spectrum \cite{he2007analyzing}. The method was extended by Weng and Lee who used wavelet analysis on word frequencies to obtain new features for every word. Based on low signal auto-correlations, trivial words were filtered out \cite{weng2011event}. Events were identified by clustering the remaining words using graph partitioning. However, these methods \cite{he2007analyzing,weng2011event} are unable to keep track of the temporal information which is a significant aspect for detecting events.

Cheng et al. modeled the word co-occurrence patterns to learn topics that resulted in inference emerging topics \cite{cheng2014btm}. Although the topic models worked well for short texts, they still required prior knowledge. Agarwal et al. discovered dense clusters in highly dynamic graphs by using a short-cycle property \cite{agarwal2012real}. These dense clusters were considered social media events. In another work, a system named DYNDENS was developed which quantified the magnitude of change based on updates in edge weights. The system incrementally computed dense subgraphs to detect event stories \cite{angel2012dense}. DYNDENS is efficient and scalable to rapidly evolving datasets. Although the detection methods proposed by \cite{agarwal2012real,angel2012dense} are efficient, despite rapid changes in microblog streams, they suffer from the loss of single-entity events. 

Traditional event detection methods are not designed to process and detect events efficiently from such dynamic data, particularly when the data stream is noisy and consists of diverse events. In addition, most of the state-of-the-art approaches depend on highly weighted and frequent patterns to detect events \cite{li2012tedas,nguyen2015real,shamma2011peaks,yang2011patterns}. These approaches ignore the dominating nature of burstiness over small events in the data.

The proposed approach differs from existing approaches because it highlights dominating patterns at an early stage in the text stream and handles post-event effects by suppressing those patterns in the subsequent time interval, which provides an opportunity to discover new emerging events. Figure \ref{fig:visualization} visualizes the pre-event, event, post-event graphs to show the characteristics of the proposed approach. Instead of focusing on burstiness, we considered change in temporal frequency with respect to time which we named displaced temporal frequency. It captured the change in the frequencies of words appearing in text stream at an early stage and later suppressed their burstiness to highlight other topics. These characteristics are an inherent part of the proposed approach, which lead to a better performance in the event detection process.

\section{Conclusion}
\label{sec:ConclusionAndFutureWork}

In this paper, we presented a novel, sensitive and efficient Weighted Dynamic Heartbeat Graph (WDHG) method to detect events from a text stream. The text stream was systematically transformed into a series of temporal graphs. These graphs inherited temporal frequencies and co-occurrence relationships of the words appearing in the text stream. Each graph was further used to extract a heartbeat score using two features: growth factor and aggregated centrality. A rule-based classifier labeled the graphs as event candidates. Multiple event candidates were merged to extract a list of ranked topics. For the performance evaluation of the proposed approach, three benchmarks: FA Cup, Super Tuesday, and the US Election were used. The quantitative evaluation showed that the proposed approach outperformed the state-of-the-art methods. The empirical evaluation showed that the proposed approach is computationally efficient and scalable. In the future, we plan to explore user participation and social network based features, as well as test the proposed approach on live text streams.

\scriptsize 

\end{document}